# Constructing a new predictive scaling formula for ITER's divertor heat-load width informed by a simulation-anchored machine learning

C.S. Chang[1], S. Ku[1], R. Hager[1], R.M. Churchill[1], J. Hughes[2], F. Köchl[3], A. Loarte[4], V. Parail[5], R. A. Pitts[4]

[1]*Princeton Plasma Physics Laboratory, Princeton University, Princeton, NJ 08540-451, USA*
[2]*Plasma Science and Fusion Center, MIT, Cambridge, MA 02139, USA*
[3]*Atominstitut, Technische Universität Wien, Stadionallee 2, 1020 Vienna, Austria*
[4]*ITER Organization, Route de Vinon sur Verdon, 13067 St Paul Lez Durance, France*
[5]*Culham Centre for Fusion Energy, Culham Science Centre, Abingdon OX14 3DB, UK*

Corresponding author's email: cschang@pppl.gov

Abstract

Understanding and predicting divertor heat-load width $\lambda_q$ is a critically important problem for an easier and more robust operation of ITER with high fusion gain. Previous predictive simulation data for $\lambda_q$ using the extreme-scale edge gyrokinetic code XGC1 in the electrostatic limit under attached divertor plasma conditions in three major US tokamaks [C.S. Chang *et al.*, Nucl. Fusion 57, 116023 (2017)] reproduced the Eich and Goldston attached-divertor formula results [formula #14 in T. Eich et al., Nucl. Fusion **53**, 093031 (2013); R.J. Goldston, Nucl. Fusion 52, 013009 (2012)], and furthermore predicted over six times wider $\lambda_q$ than the maximal Eich and Goldston formula predictions on a full-power (Q = 10) scenario ITER plasma. After adding data from further predictive simulations on a highest current JET and highest-current Alcator C-Mod, a machine learning program is used to identify a new scaling formula for $\lambda_q$ as a simple modification to the Eich formula #14, which reproduces the Eich scaling formula for the present tokamaks and which embraces the wide $\lambda_q^{XGC}$ for the full-current Q = 10 ITER plasma. The new formula is then successfully tested on three more ITER plasmas: two corresponding to long burning scenarios with Q = 5 and one at low plasma current to be explored in the initial phases of ITER operation. The new physics that gives rise to the wider $\lambda_q^{XGC}$ is identified to be the weakly-collisional, trapped-electron-mode turbulence across the magnetic separatrix, which is known to be an efficient transporter of the electron heat and mass. Electromagnetic turbulence and high-collisionality effects on the new formula are the next study topics for XGC1.

## I. Introduction

A challenge for ITER operation is the ability of the divertor plates to withstand the steady plasma exhaust heat that will be deposited on the surface along a narrow toroidal strip. A simple data-based regression using macroscopic parameters from attached-divertor experiments on all the present devices (formula #14 in Refs. [1, 2]) shows that the heat-flux width follows a scaling $1/B_{pol,MP}^{\gamma}$ where $B_{pol,MP}$ is the magnitude of the poloidal magnetic field on the outboard midplane separatrix surface and $\gamma=1.19$. References [1, 2] also present other possible regression formulas that are valid for certain chosen device sets. There has also been a heuristic model by Goldston [3] based on the neoclassical orbit-driven ion losses for weakly collisional edge plasma, which resulted in a similar result to that in [1, 2]. For ITER H-mode operation at $I_P$=15 MA with $q_{95}$ = 3, these regression and heuristic formulas yield at most $\lambda_q \approx$1mm for the divertor heat-flux width measured at outboard midplane after being mapped from the divertor plates along the magnetic field lines. Here, $\lambda_q$ is defined in the following fitting formula [1, 2]:

$h(R_{mp}-R_{mp,sep}) = 0.5 h_0 \exp[(0.5S/\lambda_q)^2 - (R_{mp}-R_{mp,sep})/\lambda_q]\ Erfc[0.5S/\lambda_q-(R_{mp}-R_{mp,sep})/S]+h_{BG}$,

where $R_{mp}$ is the major radius along the outboard midplane, $R_{mp,sep}$ is $R_{mp}$ on the outboard separatrix surface, $h(R_{mp}-R_{mp,sep})$ is the input function to the fitting formula (namely the divertor heat-flux profile data at outboard midplane after being mapped from the divertor plates along the magnetic field lines), $h_0$ is the peak value of $h$, S is a spreading parameter which makes the heat flux profile deviate from an exponential decay, *Erfc* is the complementary error function, and $h_{BG}$ is the background heat-flux.

For this range of $\lambda_q$ in ITER, the peak divertor power fluxes in attached divertor conditions are beyond the design limits of the stationary heat loads of the ITER divertor target, thus requiring the divertor operation in deeply semi-detached or detached conditions in which the plasma power is dissipated over a larger area by atomic radiation from hydrogenic-isotope atoms and impurities in the divertor chamber. The operational range for such a deeply semi-detached or detached divertor operation decreases with smaller $\lambda_q$, and is restricted to very high plasma separatrix densities and radiative fractions, requiring $n_{sep}/n_{GW}>0.6$ for $\lambda_q \approx 1$mm [4], where $n_{GW}$ is the critical plasma density inside the pedestal top above which the plasma tends to have a deteriorated confinement and even disrupt [5]. This raises concerns regarding their compatibility with the good H-mode energy confinement required to achieve Q=10 operation in ITER and the increased probability for plasma disruption. In addition, such a small $\lambda_q$ poses additional challenges for the control and sustainment of the semi-detached or detached divertor conditions since the power fluxes during transient re-attachment may significantly exceed the stationary heat flux design limits of the ITER divertor.

However, it is questionable if such a simple extrapolation from present experiments is valid as there may be differences in the fundamental edge physics between ITER and the present devices. Any extrapolation from present experiments to ITER may need to be on a more fundamental, first-principles-based kinetic physics. This was the purpose of the gyrokinetic study in Ref. [6], utilizing the edge gyrokinetic particle-in-cell code XGC1 [7].

Firstly, the heat-flux width ($\lambda_q^{XGC}$) predictions from the XGC1 gyrokinetic model reproduced the carefully chosen representative experimental data from three US tokamaks within the regression error bar of the Eich scaling study [1, 2]. Total-f gyrokinetic simulations were performed until an approximate gyrokinetic power balance was achieved in XGC1 between separatrix surface and divertor plates at the level of core heating power. A minor adjustment by the total-f XGC1 code of the experimentally measured or model profiles across the magnetic separatrix was made before approximate power balance was achieved. Secondly, the same XGC1 code was used to predict the heat-flux width on the full-current (15 MA) Q =10 ITER plasma, with the caveat that the initial ITER plasma input to XGC1 from the reduced model code JINTRAC [8] may not be in agreement with the total-f gyrokinetic code XGC1. As a matter of fact, a significant adjustment from the initial JINTRAC edge plasma happened before XGC1's achievement of an approximate gyrokinetic power-balance, between the power-crossing at separatrix and the heat load at divertor plates at the level of heat-source at the burning core.

Actual experimental plasma profiles that satisfy the Grad-Shafranov equilibrium relation required only a minor adjustment before a gyrokinetic quasi-equilibrium is reached in the total-f XGC1. However, the reduced-model predicted plasma profiles (such as those for ITER) often require a significant adjustment, in accordance with the radial plasma transport fluxes, before a gyrokinetic quasi-equilibrium is reached consistently with the magnetic equilibrium, as shown in Ref. [6] and later in the present report. There is an underlying assumption here that a deterministic gyrokinetic plasma profile state exists in accordance with external constraints when starting from different but nearby reduced-model predicted plasma profiles, as long as the external heat source profiles, the wall recycling coefficients, and the boundary conditions are identical. The most interesting finding from the study was that the same gyrokinetic code that reproduced experimental $\lambda_q$ in the present tokamak plasmas, predicted that $\lambda_q$ in the full-current ITER model-plasma in attached divertor condition would be over 6-times wider than what could be maximally extrapolated from the various Eich scaling formulas and the Goldston formula. More details can be found in Ref. [6].

Understanding the physics cause behind such a significant broadening of $\lambda_q^{XGC}$ in the full-current ITER Q = 10 edge plasma has remained as a critical research issue for the XGC group. A subsequent data analysis showed that the edge turbulence pattern across the magnetic separatrix changes from the space-time isolated "blobs" [9] in all the present tokamaks to radially extended and connected "streamers" [10] in the full-current ITER Q = 10 scenario that are typically seen in the ion-scale microturbulence such as the ion-temperature-gradient (ITG) driven turbulence and the trapped-electron-mode (TEM) turbulence. This gives us a strong hint that there is a fundamental physics change between the present tokamak edge plasma and the full-current ITER edge plasma in the XGC1 electrostatic simulation.

Another strong clue arises from the recent high-current experiments on Alcator C-Mod tokamak [11]. With the poloidal magnetic field strength as strong as that of the ITER full-current Q = 10 plasma, experimental $\lambda_q^{Exp}$ values in the Alcator C-Mod experiments still follow the Eich scaling. An XGC1 simulation has been performed on one of these high-current C-Mod plasmas and confirmed that the gyrokinetic $\lambda_q^{XGC}$ from XGC1 also follows the Eich scaling. This yields double-valued solutions for $\lambda_q^{XGC}$ between the high-current C-Mod plasma and the full-current ITER plasma if $B_{pol,MP}$ (or the macroscopic parameters used in Eich et al.) is the sole independent parameter, indicating the existence of other hidden parameter(s).

It is the purpose of the present paper to conduct a systematic search for the hidden parameter(s) and the corresponding new physics by utilizing deeper data analyses, high-fidelity physics knowledge, and a convenient machine-learning tool in search of an improved $\lambda_q^{XGC}$ scaling formula that can encompass not only all the present experimental results, but also the gyrokinetic predictions for the full-current (15MA) ITER result. Three more simulations are performed on different ITER model plasmas to successfully test the new scaling formula. The present study opens up doors to several deeper edge-physics research topics, as will be pointed out in later sections. Study of the electromagnetic and high-collisionality effects on $\lambda_q^{XGC}$ is left for future work.

We note that there is recent empirical modeling showing some widening of the near-scrape-off layer (near-SOL) upstream power-width due to a high collisionality effect [12] in present tokamaks that could represent the relative importance of the interchange effect on drift-wave turbulence [13, 14], aiming for semi-detached or detached divertor plasmas. In this work, we confine our study to the low recycling, attached divertor plasma conditions and do not attempt to study the high collisionality effect of Ref. [12]. There is a BOUT++ fluid turbulence simulation result [15] which shows broadening of $\lambda_q$ in the 15MA Q = 10 ITER plasma. Since fluid modeling does not contain the kinetic physics that are essential in the present work, such as the finite ion orbit width and trapped electron modes, we do not attempt to compare the present work with Ref. [15]. There is also a SOLPS-ITER transport modeling of the 15MA ITER discharge, with an arbitrarily chosen radial diffusion coefficient, that shows an anomalous electron thermal diffusivity at 1m$^2$/s in the SOL could broaden $\lambda_q$ to 3-4mm [16].

The paper is organized as follows: In Sec. II, for the sake of completeness, we briefly summarize the previous results from Ref. [6]. In Sec. III, we present new simulation results that answer some questions left by Ref. [6]. In Sec. IV, we utilize a machine learning program to find a new scaling formula for $\lambda_q^{XGC}$. In Sec.VI, we test the new predictive formula by performing simulations on different ITER model plasmas. In Sec. V, we describe the new physics understanding in relation to the new scaling formula. We present summary and discussion in Sec. VII.

## II. A brief summary of the previous XGC1 simulation results

In this section, for the sake of completeness, we briefly summarize the previous XGC1 simulation results reported in Ref. [6] as the basis for the discussions presented in this paper. Table I shows the seven simulation cases studied in Ref. [6], chosen in collaboration with three major US tokamaks and the ITER Organization. The discharges were selected to cover a wide range of the then experimentally available $B_{pol,MP}$, the poloidal magnetic field magnitude at the outboard midplane on the magnetic separatrix surface. Discharges from three US tokamaks were part of the discharge set used in the regression analysis in Eich et al. [1,2]. In all the discharges, the ion magnetic drift direction is toward the single magnetic X-point and the (inter-ELM) divertor plasma is in the attached regime. It should be noted here that at that time at which the work in ref. [6] was being conducted, the highest-field C-Mod experiments [11] with $B_{pol,MP}$ reaching the ITER full-current case did not exist.

| Shot | Time (ms) | $B_T$ (T) | $I_P$ (MA) | $B_{pol,MP}$ (T) |
|---|---|---|---|---|
| NSTX 132368 | 360 | 0.4 | 0.7 | 0.20 |
| DIII-D 144977 | 3103 | 2.1 | 1.0 | 0.30 |
| DIII-D 144981 | 3175 | 2.1 | 1.5 | 0.42 |
| C-Mod 1100223026 | 1091 | 5.4 | 0.5 | 0.50 |
| C-Mod 1100223012 | 1149 | 5.4 | 0.8 | 0.67 |
| C-Mod 1100223023 | 1236 | 5.4 | 0.9 | 0.81 |
| ITER full-current scenario | Flat top | 5.3 | 15 | 1.25 |

*Table I. Experimental discharges from three US tokamaks that were used by XGC in Ref. [6] and that were part of the original ``Eich-regression'' exercise [1, 2]. The last row represents a full-current, full-power ITER scenario plasma. $B_T$ is the toroidal magnetic field strength at machine axis, $I_P$ is the plasma current, and $B_{pol,MP}$ is the poloidal magnetic field strength at outboard midplane on the separatrix surface.*

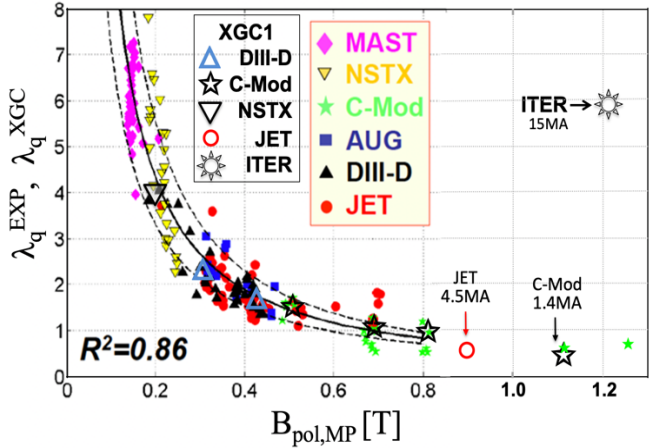

*Fig. 1. XGC that has predicted the $\lambda_q$ values in agreement with the Eich scaling formula in all three major US tokamaks predicts $\lambda_q$ = 5.9mm in a model ITER plasma edge at $I_P$=15MA, as shown in Ref. [5]. After the publication of Ref. [5], two new simulation points are added: JET at 4.5MA ($B_{pol,MP}$=0.89T, open red circle) and C-Mod at 1.4MA ($B_{pol,MP}$=1.11T, open black star) at far-right bottom. They both follow the Eich scaling formula, approximately. A couple of new high field C-Mod experimental points are also shown together (at $B_{pol,MP}$=1.1 T).*

Figure 1, without counting the 4.5MA JET and the 1.4MA C-Mod points that will be used in the next section, shows the simulation results for $\lambda_q$ from XGC1 in comparison with the experimental results $\lambda_q^{Exp}$ of Refs. [1, 2], with the symbols from XGC1 improved from Fig. 16 of Ref. [6] to resemble the corresponding experimental symbol shapes. The inaccuracy in the ITER $\lambda_q^{XGC}$=5.9mm point position in Fig. 16 of Ref. [6] is corrected in Fig. 1. As can be seen from all the open symbols, the XGC1 predictions for the present tokamaks agree well with the Eich scaling for $\lambda_q$ from formula #14 in [1] (hereafter referred to as $\lambda_q^{Eich(14)}$), represented by the solid line, together with the regression error represented by the two dashed lines. Here, we use the Eich formula #14 ($\lambda_q^{Eich(14)} \approx 0.63 B_{pol}^{-1.19}$ mm) because it contains data from all the tokamaks. Turbulence across the magnetic separatrix and in SOL was always of "blob" type in the present devices in the XGC1 simulations, as measured in some experiments. A blob is a magnetic-field-aligned intermittent plasma structure which is considerably denser than the surrounding background plasma and highly isolated in the two directions perpendicular to the equilibrium magnetic field [9]. However, the XGC1-predicted $\lambda_q^{XGC}$ in the full-current Q = 10 ITER scenario plasma (15MA, $B_{pol,mp}$=1.21T) is about 6-times greater than what could be maximally predicted from various Eich formulas/Goldston formula, or about 12 times greater than $\lambda_q^{Eich(14)}$.

In Ref. [6], a possibility for this large deviation for the full-current ITER was hypothesized to be from a much longer radial correlation length of the edge turbulence across the separatrix surface caused by the low neoclassical E×B shearing rate in the ITER full-current Q = 10 plasma. In Sec. VI, it will be shown that the turbulence with much longer radial-correlation length has a streamer structure, which is usually observed in ITG and TEM driven turbulence. This hypothesis was drawn from the fact that the neoclassical physics strength, thus the neoclassical E×B flow shearing rate, becomes weaker as $\rho_{i,pol}/a$ becomes smaller, where "$\rho_{i,pol}$" is the poloidal ion Larmor radius at the outboard midplane separatrix point and "$a$" is the plasma minor radius. In the full-current ITER, $\rho_{i,pol}/a$ is an order of magnitude smaller than that in the highest-current C-Mod plasma. In the present tokamak devices, XGC1 found that the divertor heat-flux width physics is dominated by the ion neoclassical drift motions [6], in agreement with Ref. [3], in spite of the existence of large-amplitude blobby turbulence across the separatrix and in the SOL.

A quick demonstration of the neoclassical E×B dependence on ion banana width can be given by using the standard neoclassical radial force balance equation in the closed field-line region [17]:

$$\langle u_\| \rangle + (T_i/neB_p)dn/dr = (1/eB_p)[(k-1)dT_i/dr - ed\langle \Phi \rangle/dr], \quad (1)$$

where $\langle u_\| \rangle$ is the flux-surface averaged parallel fluid-flow velocity and k is a collisionality-dependent parameter that is 1.17 when ions are in the banana regime [17] (ions near the magnetic separatrix in the full-current ITER edge are in this regime, but the value k=1.17 may not be accurate in the edge plasma). Neglecting, for the sake of a simpler argument, the temperature gradient term, whose gradient and coefficient are significantly smaller than the density gradient term for k=1.17, we can simplify and rearrange Eq. (1) to

$$u_E/v_{i,pol} - \langle u_\| \rangle/v_i \approx \rho_{i,pol}/\alpha a, \quad (2)$$

where $u_E = E_r/B$ is the E×B flow speed, $v_i$ is the ion thermal speed, $v_{i,pol}$ is the poloidal component of the parallel thermal speed, $\rho_{i,pol}$ is the ion gyroradius in the poloidal magnetic field, and $\alpha a$ is an expression for the density gradient scale length expressed in terms of a parameter $\alpha$ and the plasma minor radius. For H-mode pedestals in the conventional aspect-ratio tokamak edge, $\alpha$ does not vary widely but stays around ~0.05. It can be easily noticed from Eq. (2) that the plasma gradient term $\rho_{i,pol}/\alpha a$ is the driver for the radial electric field, or equivalently for the E×B flow that is mostly in the poloidal direction. As the device size becomes greater relative to the ion poloidal gyroradius, $u_E$ becomes smaller in proportion. For the full current 15MA ITER, $\rho_{i,pol}/a$ is about 6 times smaller than the 1.5MA DIII-D case of Table I. The physically meaningful E×B shearing parameter is $\gamma_E/\omega_* = du_E/(\omega_* dr)$ which scales as, using Eq. (2) and the relations $\omega_* \sim kv_i\rho_i/\alpha a$, $du_E/dr \sim u_E/L_E$, and assuming $k\rho_{i,pol} \sim 1$,

$$\gamma_E/\omega_* \sim (du_E/v_i\, dr)(\rho_{i,pol}/\rho_i)\alpha a \sim v_{i,pol}(\rho_{i,pol}/\alpha a L_E)(\rho_{i,pol}/\rho_i)\alpha a \sim \rho_{i,pol}/L_E. \quad (3)$$

This relationship shows that $\gamma_E/\omega_*$ decreases with $\rho_{i,pol}/L_E$ ($\propto \rho_{ip}/a$). Value of the XGC-found $\gamma_E/\omega_*$ for representative device cases will be shown in Sec. VI, where detailed physics is discussed.

III. New XGC1 simulations

The XGC family codes are equipped with a built-in Monte Carlo neutral particle transport capability using ionization and charge exchange cross-sections for neutral-plasma interaction. A recycling coefficient R=0.99 is used for the divertor heat-load width simulations presented here, for generation of neutral marker-particles at Frank-Condon energy (3eV) in front of the material wall wherever the ions are absorbed. For a more detailed introduction, we refer the reader to Ref. [18]. In addition to the built-in Monte Carlo neutral particle transport routine, XGC family codes can utilize the DEGAS2 Monte Carlo neutral particle code as a subroutine, which can start the neutral particle recycling process from molecular neutral birth, with volumetric and surface recombination. The latter features are not utilized in the present simulations; hence our study is limited to the attached, low-recycling divertor regime. We also use a simple cooling profile in the divertor chamber to keep the electron temperature on the outboard separatrix surface close to the input value.

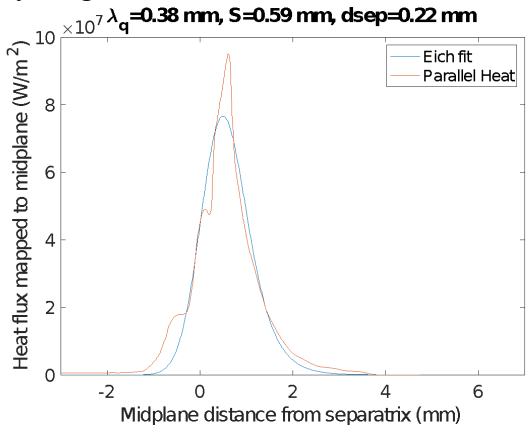

*Fig. 2. Footprint of the parallel heat-flux on the outer divertor plates, mapped to the outboard midplane, in one of the highest current C-Mod discharges (discharge number #1160930033).*

The first new XGC1 simulation is to test an existing experimental plasma that is closest to the full-current ITER in both the $B_{pol,MP}$ value and the physical size in deuterium plasma. For this purpose, a JET 4.5MA discharge [19] is chosen that has the highest $B_{pol,MP}$ (0.89T) at the time of simulation (unfortunately, an experimental $\lambda_q^{Exp}$ measurement does not exist on JET at this high value $B_{pol,MP}$). To be more specific, $B_{pol,MP}$ for this JET plasma is only 36% lower than the full-current ITER plasma, and its linear size is a factor of $\approx 2$ smaller than ITER. In this plasma, XGC1 finds $\lambda_q^{XGC}$ of about 0.64mm, which is within the regression error bar from the Eich(14) value $\lambda_q^{Eich} \cong 0.72$mm (open red circle in Fig. 1). Thus, XGC1 indicates that there may be either a bifurcation of $\lambda_q^{XGC}$ between $B_{pol,MP} = 0.89$T of JET and 1.21T of ITER, or there is something other than the value of $B_{pol,MP}$ which sets the full-current ITER case apart from the present experimental scaling.

Shortly after the JET simulation described above was performed, experiments at C-Mod raised $B_{pol,MP}$ values up to 1.3T [11] which somewhat exceeds the full-current ITER value, and found that the experimental $\lambda_q^{Exp}$ still follows $\lambda_q^{Eich}$ approximately. This was an excellent comparison case to be studied by XGC1. Accordingly, we chose the C-Mod discharge #1160930033 with 1.4MA of plasma current and $B_{pol,MP} = 1.11$T. At this high value of $B_{pol,MP}$, though, we find $\lambda_q^{XGC} \cong 0.38$mm (see fig. 2 and the open black star symbol at the far-right bottom of Fig. 1), which is even somewhat smaller than $\lambda_q^{Eich(14)} = 0.56$mm. As a result, XGC1's solution becomes double valued around the maximal C-Mod $B_{pol,MP}$ values if $B_{pol,MP}$ is used as the sole parameter, and suggests existence of hidden parameter(s) that was missed in Eich's regression process.

IV. A simulation-anchored, predictive machine learning study

In this section, we use a supervised machine learning program in search of the possible hidden parameter(s). A machine learning program is basically a systematic interpolation and regression technique utilizing mathematical tools. A machine-learning program can yield answers much more rapidly and systematically than human interaction with ordinary spreadsheets can. Any presently available data set forms an underdetermined system, which is only a subset of all the possible data sets and which may not be good for extrapolation into a new regime where the governing physics phenomena may be different. An extrapolation path from the present data knowledge alone could lead us to a completely wrong direction. However, if a first-principles model can be used to study the new regime and make predictions in accordance with the new governing physics, the simulation results can "anchor" the machine learning into the new physics direction, at least as far as the simulation correctness in the specific target regime is concerned. The "anchoring" high-fidelity simulation points do not have to be many to lead the machine learning prediction into the intended direction: but, the more the better for accuracy. Of course, the accuracy of the simulation-anchored predictive machine learning will only be as good as the anchoring high-fidelity model accuracy, which will improve as the computational power increases (or a high-fidelity analytic model). We caution here that the simulation must be well-validated on the present experimental data before adding the anchoring data. The extrapolated predictions must also be validated continuously against new experiments when available.

In this section, we use this "anchored machine learning" concept to search for a predictive analytic scaling formula by combining the experimental and predictive-simulation data sets for the divertor heat-flux width $\lambda_q$. We use the symbol $\mathbf{D}^E$ to represent a set of $\lambda_q^{Exp}$ data found from the present laboratory experimental measurements, $\mathbf{D}^{SE}$ for a set of $\lambda_q$ data found through high-fidelity simulation of the existing experiments, and $\mathbf{D}^{SF}$ for a set of $\lambda_q$ data found through high-fidelity simulation of future experiments. We use $\mathbf{M}$ to denote the machine-learning operation, $\mathbf{F}^E$ for the modeling formula found by the operation $\mathbf{M}$ on the present experimental data set $\mathbf{D}^E$, $\mathbf{F}^{SE}$ for the modeling formula found by the operation $\mathbf{M}$ on $\mathbf{D}^{SE}$, and $\mathbf{F}^P$ for the predictive modeling formula found by the operation $\mathbf{M}$ on all the data sets including $\mathbf{D}^E$, $\mathbf{D}^{SE}$, and $\mathbf{D}^{SF}$. $\mathbf{D}^E$ and $\mathbf{D}^{SE}$ do not need to have one-to-one correspondents.

For the validated high-fidelity simulations, we assume $\mathbf{F}^E \approx \mathbf{F}^{SE}$ as a pre-requisite condition, which is satisfied by XGC1 as shown in the previous sections. Thus, we have $\mathbf{M}(\mathbf{D}^E) \to \mathbf{F}^E$ and $\mathbf{M}(\mathbf{D}^{SE}) \to \mathbf{F}^E$, with some allowance for error. We can then write down the following relations

$$\mathbf{M}(\mathbf{D}^E \cup \mathbf{D}^{SE}) \to \mathbf{F}^E, \text{ and} \tag{4}$$

$$\mathbf{M}(\mathbf{D}^E \cup \mathbf{D}^{SE} \cup \mathbf{D}^{SF}) \to \mathbf{F}^P[\supset \mathbf{F}^E]. \tag{5}$$

Here, $\mathbf{F}^P[\supset \mathbf{F}^E]$ means that the machine-learned formula $\mathbf{F}^P$ reduces to $\mathbf{F}^E$ in the present-day experimental space. In other words, using predictions from simulation on the unexplored future experiments, the simulation-anchored machine-learning operation can be made to possess the predictive capability $\mathbf{F}^P$, within the simulation accuracy, by combining $\mathbf{D}^E$ and $\mathbf{D}^{SE}$ with $\mathbf{D}^{SF}$.

To achieve this goal, we use an AI-based modeling engine Eureqa [20, 21]. Eureqa uses supervised machine learning techniques to conduct an evolutionary model search to find the best combination of the user-specified mathematical building blocks that fit labeled training data, not only equation parameters, but also the form of the symbolic equation which best fits the data [22]. Starting with a series of random expressions, the algorithm combines the best-fitting expressions with each other until it gradually finds formulas which fit the data. Eureqa also applies a penalty in proportion to the complexity of the formula so as to prevent overfitting. While trial-and-error single fits could be performed using different forms of equations on combinations of parameters, using symbolic regression frees us from specifying the form of equations to fit the data, resulting in more generic equations.

Our attempt is to find a new predictive scaling formula $\mathbf{F}^P$ of Eq.(4). We present the result first: Fig. 3 depicts the simplest $\mathbf{F}^P$ result from Eureka, as to be elaborated soon later in this section. Figure 3 contains the selected experimental data set $\mathbf{D}^E$ from NSTX, DIII-D, and C-Mod (marked with + symbols) as presented in Sec. II, and the corresponding simulation data set $\mathbf{D}^{SE}$. The purely predictive 4.5MA JET and 15MA ITER simulations, for which experimental measurements do not exist, are also contained Fig. 3. We have normalized all the $\lambda_q$ values in $\mathbf{D}$ of Eqs. (4) and (5) to the Eich scaling formula #14, $\lambda_q^{Eich(14)} = 0.63\, B_{pol,MP}^{-1.19}$. The simple extrapolation to the future experiments from the present-day experimental data set is represented by the solid black horizontal line.

Observables in tokamak plasmas are functions of many variables and the machine learning can be a many-variable operation. Eich *et al.* used the nine well-known macroscopic variables for a thorough data regression [1, 2], which spans the macroscopic plasma-operation space rather completely: $B_{tor}$ (the toroidal magnetic field strength), $B_{pol,MP}$, $q_{95}$ (the safety factor at the 95% poloidal-flux surface), $P_{SOL}$ (the power flow from core into the SOL), $R_{geo}$ (the geometric major radius), $a$ (the plasma minor radius), $I_p$ (the plasma current), and $n/n_{GW}$ (the density ratio to the Greenwald density). Multiple possible formulations are found from data regression in Refs. [1, 2] depending on the combination of the target tokamaks, but the main dependence of the divertor heat-flux width is found to be on $B_{pol,MP}$ by targeting all the present tokamaks, denoted here as the Eich regression number #14, with the squared correlation coefficient being $R^2=0.86$. Our machine learning operation utilizes $\lambda_q^{Eich(14)}$ as the normalization factor.

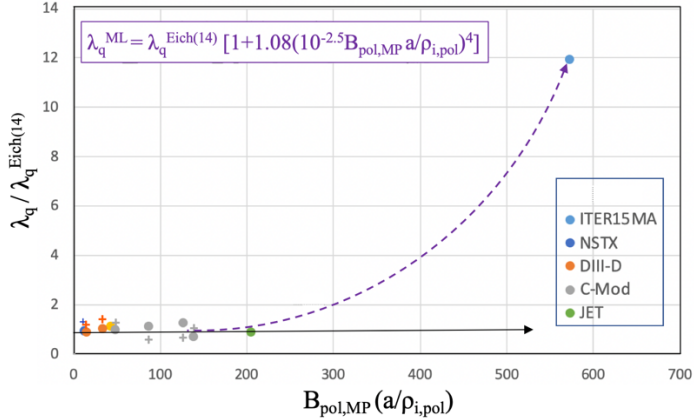

*Fig. 3. A new $\lambda_q$ formula (purple equation and dashed line) from simulation-anchored machine learning: Among the color-filled circles, NSTX (dark blue), DIII-D (orange) and C-Mod (gray) represent the XGC1 predicted data set $\mathbf{D}^{SE}$ for the experiments that has experimental data set $\mathbf{D}^E$, marked with + sign with the same coloring scheme. The JET and the 15MA ITER data points are the simulation data set $\mathbf{D}^{SF}$ that do not have experimental measurements. The solid black arrow shows a simple extrapolation from the present experiments to the unknown physics regime.*

We note that Refs. [1] and [2] did not consider microscopic kinetic parameters. Among the microscopic kinetic parameters, there is a dimensionless quantity that could be as important as the macroscopic parameters: the ratio between the ion banana width to the device size [6, 23] as elaborated at the end of Sec. II. The ratio between the ion banana width and the machine size determines the strength of neoclassical physics [see Eq. (2)], including the important background $E_r \times B$-flow shearing rate (see Eq. (3)) which controls plasma turbulence [24]. Plasma turbulence could then affect the cross-field spread of the divertor heat-load (characterized by $\lambda_q$). For this reason, we introduce a new parameter "$a/\rho_{i,pol}$" to be used for a physics-based feature-engineering in the supervised machine-learning in Eureqa. Comparison of the normalized $E \times B$-flow shearing rate for example tokamaks that have different $a/\rho_{i,pol}$ values will be presented in section VI.

Our first try in the present work is to accept the regression result of Refs. [1, 2], thus accept that there is little dependence of $\lambda_q$ on all other macroscopic parameters, and utilize only two parameters in the machine learning program Eureqa: $B_{pol,MP}$ inherited from Refs. [1, 2] and the kinetic parameter $a/\rho_{i,pol}$. If this simplified approach does not work satisfying our three conditions – to resolve the double valued solution issue, to agree with the well-validated $\lambda_q^{Eich}$ formula for the present attached divertor experiments, i.e., $\mathbf{F}^P[\supset \mathbf{F}^E]$, and to encompass the full-current ITER Q = 10 result – then we will have to ignore the work done in Refs. [1, 2] and perform a many variable machine learning study from scratch.

Application of the data sets $\mathbf{D}^E \cup \mathbf{D}^{SE} \cup \mathbf{D}^{SF}$ to Eureqa then gave us numerous possible predictive modeling formulas, most of which turn out to be some complicated and meaningless functional combinations of the input parameters $B_{pol,MP}$ and $a/\rho_{i,pol}$. Three physics-based search-formulas are given to Eureqa to shorten the search time to one hour on a MacBook Pro equipped with a 2.6 GHz Intel Core i7 4-core processor:

$$\lambda_q/\lambda_q^{Eich\,(14)} = f(B_{pol,MP},\ a/\rho_{i,pol}),$$
$$= f(B_{pol,MP},\ a/\rho_{i,pol},\ B_{pol,MP}\ a/\rho_{i,pol}),\ \text{and}$$
$$= f(B_{pol,MP}\ a/\rho_{i,pol}).$$

Among the simulation-anchored formulas found by Eureqa,

$$\lambda_q^{ML} = 0.63 B_{pol,MP}^{-1.19} [1.0 + 1.08 \times 10^{-10} (B_{pol,MP} a/\rho_{i,pol})^4] \text{ with RMS error} = 18.7\% \quad (6)$$

is the simplest and lowest order expression for the heat flux width $\lambda_q^{ML}$ derived by this Machine Learning approach with a reasonably low mean square error (RMS error = RMSE = 18.7%). Eq. (6) is depicted in Fig. 3 using the dashed purple curve. A lower-order formula could not be picked because the mean square error jumps to above 50%. The formula agrees with $\lambda_q^{XGC}$ for the full-current ITER plasma and reproduces $\lambda_q^{Eich(14)}$ for all the present-day tokamak data. The predictive simulation on the 4.5MA JET plasma (for which the experimental data does not yet exist) contributes valuably to the 4$^{th}$ power law in the $B_{pol,MP} a/\rho_{i,pol}$ dependence. Notice here that in Fig. 1, the right-most data point used for the XGC1 simulation is from the high field C-Mod. In Fig. 3, however, the right-most data point became the JET simulation point indicating that the highest-field JET case is the closest present tokamak device to the full-current 15MA ITER as far as $\lambda_q$ is concerned in this parameter space.

Other example formulas that utilize combinations of $B_{pol,MP}$ and $a/\rho_{i,pol}$, and that yield low RMSE include

$$\lambda_q^{ML} = 0.63 B_{pol,MP}^{-1.19} [1.0 + 1.961 \times 10^{-16} (a/\rho_{i,pol})^6], \text{RMSE} = 17.9\% \quad (7)$$

$$= 0.63 B_{pol,MP}^{-1.19} [1.0 + 1.68 \times 10^{-18} B_{pol,MP} (a/\rho_{i,pol})^7], \text{RMSE} = 17.0\% \quad (8)$$

$$= 0.63 B_{pol,MP}^{-1.19} [1.0 + 1.91 \times 10^{-13} (B_{pol,MP} a/\rho_{i,pol})^5], \text{RMSE} = 17.0\% \quad (9)$$

$$= 0.63 B_{pol,MP}^{-1.19} [1.0 + 3.46 \times 10^{-4} [4.04 \times 10^{-5} B_{pol,MP} (a/\rho_{i,pol})^3]^{B_{pol,MP}}], \text{RMSE} = 16.1\%. \quad (10)$$

All these formulas yield fitting curves that have similar levels of RMSE to Eq. (6), matching the $\lambda_q$ values for the existing tokamaks and the "anchored" full-current ITER as well as Eq. (6) does. However, they have higher order and/or more complicated parameter dependencies, which could make the fitting curve behave differently in the gap region between the present tokamaks and the full-current ITER. In the next section, we test Eq. (6)-(10) by performing XGC simulations on three more ITER model plasmas. The results do not suggest that we should switch Eq. (6) to a more complicated formula. Besides Eqs.(6)-(10), there are other highly complicated and non-smooth formulas Eureqa has produced that try to fit details of the noisy data with much lower mean-squared error (as low as RMSE~4.5%). However, theses formulas do not reproduce the smooth Eich experimental formula and do not satisfy the requirement to reproduce the Eich regression #14 formula.

A schematic diagram for the workflow used to find the above machine-learned formulas is depicted in Fig. 4, showing the inputs (labeled experimental and simulation data for $\lambda_q$, $B_{pol,MP}$, $a/\rho_i$; mathematical operations; and variables), the evolutionary model search process in Eureqa, and the resulting $\lambda_q^{ML}$ formulas (only one of them is shown).

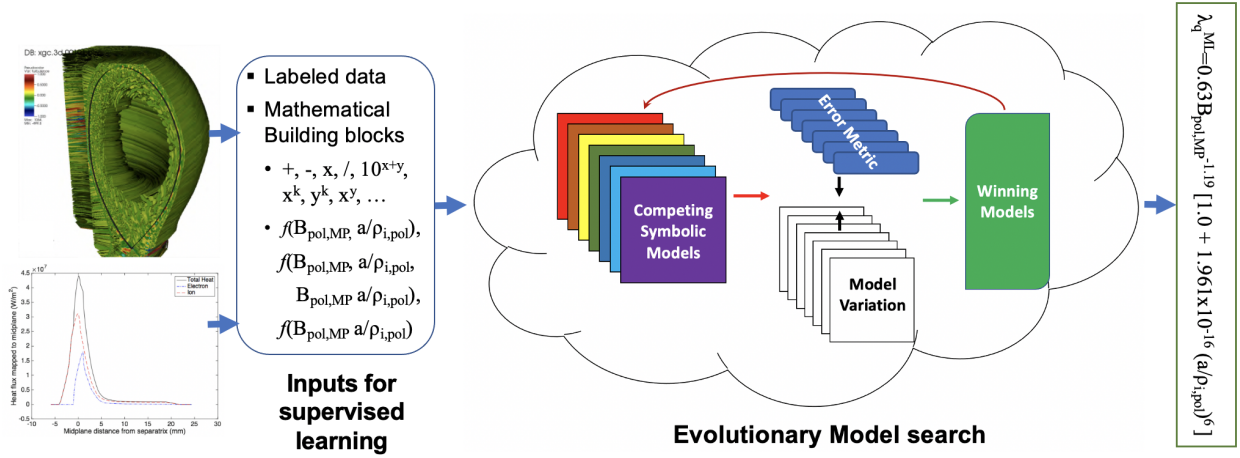

Fig. 4. Workflow used in the present supervised machine-learning study: The evolutionary model search is performed by the Eureqa program [20, 21].

V. Test of the new formula

The new ML-found formula is tested on three different ITER model plasmas: i) the first H-mode plasma to be explored in the initial phases of ITER operation at 5MA [25], ii) an H-mode hybrid plasma at 12.5MA providing long pulse operation with fusion yield Q = 5 [26], and iii) an H-mode steady-state plasma at 10 MA providing steady-state operation with Q = 5 [27]. These three ITER model plasmas have distinctively different values of the kinetic parameter $a/\rho_{i,pol}$ at the outboard midplane edge. The 5MA plasma has $a/\rho_{i,pol}$ that is well within the present tokamak range, but its physical size is the same as the full-current ITER plasma; the 12.5MA hybrid plasma has $a/\rho_{i,pol}$ slightly above the 15MA plasma and is thus a good test problem to confirm/refute the large $a/\rho_{i,pol}$ effect found on the 15MA plasma; and the 10MA steady-state plasma has $a/\rho_{i,pol}$ deep in the gap region between the high-field JET plasma and the 15MA ITER plasma. As for the original 15 MA Q=10 ITER discharge, all the new ITER points assume deuteron plasma only and do not include impurity species, but with realistic electron mass. For a visual introduction, results from three new cases are depicted in Fig. 5, as additions to Fig. 3, before being described below in more detail.

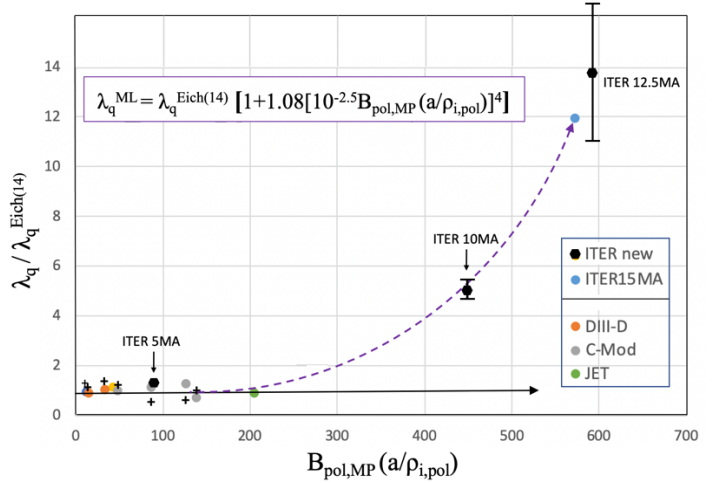

Fig. 5. Addition of three new ITER test cases to Fig. 3. The errorbar is relatively large at large $B_{pol,MP}\ a/\rho_{i,pol}$ (20%, at the ITER 12.5MA point) compared to the small $B_{pol,MP}\ a/\rho_{i,pol}$ cases (6.3%, at the ITER 10MA point).

We note here that an extension of the high current (4.5MA) JET plasma that is modeled toward the $B_{pol,MP}\ a/\rho_{i,pol}$ value of the 15MA ITER discharge could have been an option instead of the 10MA ITER case. The plasma equilibrium has to be made up in both cases, which would certainly not be in gyrokinetic equilibrium and must be evolved significantly by XGC1 before power balance between the separatrix and the divertor plates is reached. We choose the 10MA ITER case here because of the relevance of the 10MA ITER H-mode scenario for steady-state demonstration at Q=5. Our simulation can be taken as a gyrokinetic base for predictions of a future real experiment that is planned to be executed and that can be compared with future SOLPS-ITER simulations for these plasmas. A JET experiment at much higher plasma current than 4.5MA in the present divertor geometry is beyond the capabilities of the device and thus cannot be realized (nor will it be simulated by fluid codes).

i) 5MA ITER case

After the previous XGC publication of the significantly enhanced divertor heat-flux width in the ITER full-current scenario plasma [6], a question naturally arose if the enhancement could simply be from the pure size-effect: ITER is about 3-times as large as DIII-D and 9-times as large as Alcator C-Mod in linear size, with its plasma volume approximately $3^3$- and $9^3$-times greater. The first H-mode plasma scenario that will be explored in the initial ITER experimental phases with $I_p$=5MA [25] is an excellent case to answer this question: It has $B_{pol,MP}$=0.43T, similar to a high-field DIII-D plasma and a low-field C-Mod plasma (see Table I and Fig. 1), while the plasma size essentially the same as the full-current ITER. The $a/\rho_{i,pol}$ value of 201 is also similar to a typical JET plasma value, with our new parameter $B_{pol,MP}\,a/\rho_{i,pol}$ for 5MA ITER falling well within the present device range (see Fig. 5). For a quantitative comparison, the $B_{pol,MP}\,a/\rho_{i,pol}$ value for the 5MA ITER case is as small as 87, with $B_{pol,MP}\,a/\rho_{i,pol}$ for all the present tokamak experiments falling between about 10 and 200. The test XGC1 simulation finds $\lambda_q^{XGC}$ = 2.2mm, which satisfies the Eich formula value $\lambda_q^{Eich(14)}$=1.7mm approximately within the regression error bar. This result thus excludes the pure size effect from the possible cause for the large $\lambda_q^{XGC}$ found for the full-current 15 MA Q = 10 ITER plasma.

ii) 12.5MA Q = 5 long pulse ITER hybrid scenario case

The 12.5MA ITER hybrid scenario plasma with $B_{tor}$=5.3T and fusion gain of Q=5 [26] is an interesting case. Its toroidal magnetic field strength at the machine axis $B_{tor}$=5.3T is the same as the full current 15MA case. However, because of the stronger Shafranov shift due to the higher beta and a somewhat smaller major radius of outer-midplane separatrix, the value of $B_{pol,MP}$ (=1.22T) for the 12.5MA case is about the same as that (1.21T) in the 15MA discharge. Due to the smaller ion temperature at the edge (we use plasma values at $\psi_N$=0.99), the new parameter $B_{pol,MP}\,a/\rho_{i,pol}$ for the 12.5MA case is actually slightly greater than the 15MA case (592T versus 572T). This is an interesting case that may be at odds with conventional ITER H-mode plasmas between 5MA and 15MA (with similar beta and H98 = 1) in the $a/\rho_{i,pol}$ kinetic parameter space, but an excellent second case for testing the broadening of $\lambda_q^{XGC}$ by the large $B_{pol,MP}\,a/\rho_{i,pol}$ effect. A peculiarity of this plasma scenario will appear again in the discussion on the in-out asymmetry of the divertor power load in VI. Our simulation shows that $\lambda_q^{XGC}\approx$6.9mm for this 12.5MA ITER model plasma, as depicted in Fig. 5. This value is indeed somewhat greater than $\lambda_q^{XGC}\approx$5.9mm found on the full-current ITER model plasma, consistently with a slightly greater $B_{pol,MP}\,a/\rho_{i,pol}$ value. Thus, our new formula passes this test, too.

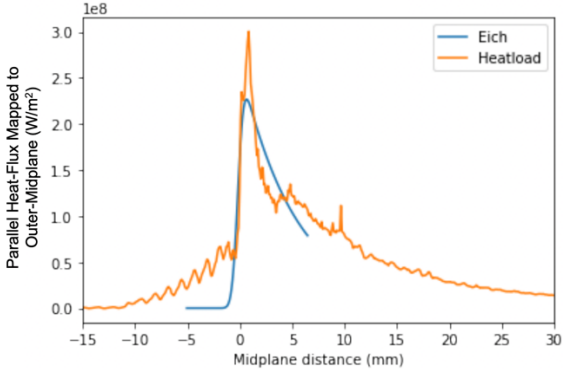 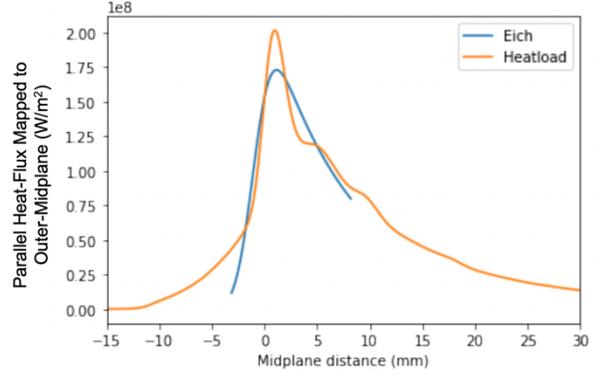

*Fig. 6. Eich formula fitting over the raw XGC1 data, mapped to the outer midplane, for the 12.5MA ITER case to obtain $\lambda_q^{XGC}$=5.5mm. To emphasize the heat-load at the peak, the fit is performed using data between -5mm and +6mm. The blue fitting curve is cropped to show the fitting data range. The long tail into far SOL is ignored.*

*Fig. 7. Eich formula fitting over the 9-point (Δr~0.8 mm) moving averaged XGC1 data along the divertor plates for the 12.5MA ITER case. The same fitting range is used as in Fig.6. $\lambda_q^{XGC}$ = 8.2mm is obtained from this fitting. Again, the blue curve is cropped to show range of the data used for fitting.*

At this point, we mention the error/uncertainty range in the Eich-formula fitting of the XGC1 data for the 12.5MA ITER case. The Eich fitting formula, as described in Refs. [1, 2], itself is well defined. The uncertainty range of $\lambda_q^{XGC}$ fitting for the present devices was smaller than the Eich regression error range and was not discussed in Ref. [6] (the ITER 10MA case can be used as an example, to be presented later in this section). However, at such a large $\lambda_q^{XGC}$ as in the 12.5MA ITER case, we find that the noisy fluctuations in the heat-flux footprint are surfacing in the raw simulation data due to the smallness of the radial resolution compared with $\lambda_q^{XGC}$ (see Fig. 6). This type of fluctuation in the XGC footprint is most likely from numerical noise due to particle noise, and may not represent what is seen in the experiment. Possible difference between the numerical heat-flux measurement and the experimental thermal sensor measurement is the reason why we call $\lambda_q^{XGC}$ the "heat-flux" width instead of the "heat-load" width. A long tail into the far scrape-off layer (SOL) can be noticed, which is unimportant for the peak divertor heat-load density. We can smooth out the footprint until the noisy fluctuation disappears. This introduces arbitrariness and uncertainty in the $\lambda_q^{XGC}$ value measurement.

In the 12.5MA ITER case, the raw data gives the narrowest $\lambda_q^{XGC}$(min) fitting due to the sharp peak near the separatrix leg (see Fig. 6), caused by the parallel electron heat flow. In our Eich-formula fitting of $\lambda_q^{XGC}$, we try to emphasize the peak heat-load density around the separatrix leg. We find $\lambda_q^{XGC}$(min)=5.5mm. We then smooth the footprint data until all the noisy fluctuations disappear before estimating the widest possible $\lambda_q^{XGC}$(max). Here we apply a 9-point (Δr~0.8 mm) moving-averaging in the radial direction and obtain $\lambda_q^{XGC}$(max)= 8.2mm (see Fig. 7). The point depicted in Fig. 5 is the midpoint between these two values, with the error bar of about $\pm 20\%$ calculated from the maximal and minimal $\lambda_q^{XGC}$ values. This type of uncertainty analysis was not performed on the 15MA case in Ref. [6], but it can be assumed that a similar level of uncertainty exists.

iii) 10MA Q = 5 steady-state ITER scenario case

There is a wide gap in the new parameter ($B_{pol,MP} a/\rho_{i,pol}$) space between the high-current JET plasma and the 15MA ITER plasma. To check the validity and accuracy of the new machine-learned $\lambda_q^{ML}$ formula, it is necessary to have at least one predictive simulation deep in the gap region as explained earlier. For this purpose, we pick the 10 MA Q = 5 ITER steady-state model plasma (see Fig. 5). XGC1 finds that $\lambda_q^{XGC}$ from the raw footprint data is 2.5mm and from the smoothed data is 2.8mm. If we take 2.5mm as the theoretical minimum value and 2.8mm as the theoretical maximum value, the median value 2.65mm and the error bar ($\pm 6\%$) are marked in Fig. 5. The difference in the $\lambda_q^{XGC}$ fitting between the raw data and the smoothed data is not as great as the 12.5MA case since the finite radial grid size has already provided some smoothing (given that the spreading is lower than at 12.5 MA). As can be seen from Fig. 5, the validity of the new simple formula is remarkably good.

Since the 10MA ITER case is located deep in the gap between 4.5MA JET and 15MA ITER, this is a good case to check the consistency of the formulas Eqs. (6)-(10) with the $\lambda_q^{XGC}$=2.65mm value found from Eq. (6). The following table summarizes the comparison. For reference, $\lambda_q^{Eich(14)}$=0.53mm. It can be seen that the simplest formula, Eq. (6), is the most consistent one with the XGC-found $\lambda_q^{XGC}$ value for the 10MA ITER case.

| Formula No. | $\lambda_q^{ML}$ from various formulas | Ratio to $\lambda_q^{XGC}$=2.65mm |
|---|---|---|
| Eq. (7) | 2.77 mm | 1.05 |
| Eq. (8) | 0.86 mm | 0.32 |
| Eq. (9) | 1.79 mm | 0.68 |
| Eq. (10) | 2.30 mm | 0.87 |
| Eq. (11) | 2.24 mm | 0.84 |

*Table II. Ratio of the $\lambda_q^{ML}$ values from various ML-found formulas to the XGC-found $\lambda_q^{XGC}$(=2.65mm) for the 10MA ITER scenario plasma where $a/\rho_{i,pol}$ = 384 and $B_{pol,MP} a/\rho_{i,pol}$ = 446 (T).*

VI. New physics understanding and its relevance to the predictive formula

As explained in Sec. II, the new parameter "$a/\rho_{i,pol}$," representing the ratio between the device size and the ion poloidal gyroradius ($\approx$ ion banana width in the edge plasma) comes from the important kinetic micro-physics that was not part of the macro-parameter set used in Refs. [1, 2]. This ratio determines the strength and weakness of the neoclassical effects, which include the background E×B-flow shearing rate (see Eq. (3)) that can control plasma turbulence [24]. As the "$a/\rho_{i,pol}$" ratio becomes higher, the neoclassical E×B-flow shearing effect gets weaker, turbulence modes that were otherwise suppressed by a strong shear-flow could surface and, at the same time, the E×B-shear-flow driven turbulence can recede.

To investigate if there is a physics difference between the full-current ITER edge and the tokamak edge that follows the Eich/Goldston-scaling, we compare the turbulence property between the full-current 15 MA ITER edge that has much greater $a/\rho_{i,pol}$ than in today's tokamaks than that of the 5MA ITER edge. This choice removes the pure, absolute size effect in the comparison. Figure 8 depicts a snapshot pattern of the normalized electron density fluctuation $\delta n/n$ obtained from the XGC1 simulations around the outboard midplane across the magnetic separatrix surface (vertical dashed line). It can be seen that across the outboard separatrix surface of the 5MA ITER H-mode plasma, plasma turbulence is of the isolated blob type as seen in both XGC1 simulations and laboratory experiments on today's tokamaks [9]. However, in the zoomed-in figure for the 15MA full-current ITER, the turbulence is of radially extended/connected streamer type as usually seen in ITG and TEM turbulence [10].

For a deeper understanding of the turbulence modes, we study the phase correlation between the electron density fluctuation δn and the electrostatic potential fluctuation δΦ, and plot them in Fig. 9. When the electrons behave adiabatically, which is a typical signature of TEM modes, the phase correlation vanishes and the raidal transport vanishes. It can be easily noticed that the electrons in the near-SOL have small phase correlation coefficient between δn and δΦ, hence are more adiabatic in the 5MA ITER edge, which is the region where the $\lambda_q^{XGC}$ footprint is measured, while they are strongly non-adiabatic in the near-SOL of the 15MA ITER edge – actually, the strongly non-adiabatic region starts just inside the separatrix into the near-SOL. This is indication that the streamer type fluctuations seen in the 15MA ITER have a strong TEM component. An ITG dominant turbulence has a stronger adiabatic electron response.

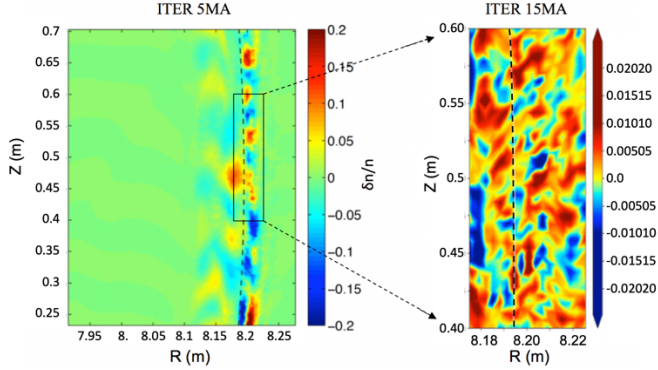
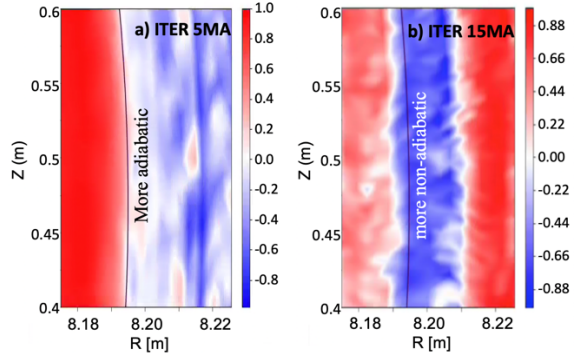

*Fig. 8. Comparison of the edge turbulence pattern in δn/n obtained from XGC1 between 5MA ITER and 15MA ITER across the separatrix surface. Isolated blob structure can be seen in the 5MA case, with the relative density fluctuation amplitude becoming large across the separatrix. On the other hand, turbulence becomes a connected streamer type in the 15MA case.*

*Fig. 9. Comparison of the phase correlation between the electron density fluctuation δn and the electrostatic potential fluctuation δΦ around the outboard midplane edge. It can be noticed that the electrons are nearly adiabatic in the near-SOL of the 5MA ITER edge, while they are strongly non-adiabatic in the 15MA ITER edge.*

The third data analysis we performed is a simple unsupervised machine-learning analysis of the electron-response correlation to the edge turbulence just outside of the separatrix surface [28]. The K-Means Clustering method in APACHE Spark [29] is used to divide the electron response into six groups with each group represented by different colors. The result is depicted in Fig. 10 as a contour plot in two-dimensional velocity space (reprint from Fig. 3 of Ref. [28]). It can be seen that the electrons are grouped mostly in energy – a sign of kinetic-energy dependent oscillations – except around $(v_\parallel^2 + v_\perp^2)^{1/2} \sim 2$ where there is a distinctively different response between the trapped and passing electrons. In this energy band, dark navy blue and medium sapphire blue are separated at the trapped-passing boundary. This is a sign of trapped electron mode driven turbulence. Different behavior around $v_\parallel \sim 0$ in the trapped electron response band is not a surprise since the deeply trapped electrons around the outboard midplane do not experience much toroidal precession drift (TEMs are driven by resonance between toroidal precession drift of the trapped electrons and drift waves). Higher number of clustering groups could show a more detailed and gradual change. The vertical Landau resonance pattern in accordance with $k_{\parallel}v_{\parallel}\sim\omega$ is not seen, indicating that the turbulence may not be from ITG modes. Besides, there is an evidence in the literature that ITG modes cannot survive in the SOL [30].

All three pieces evidence (streamer-like structures, non-adiabatic electrons, and different response of trapped electrons from passing electrons at a specific energy band) suggest that the turbulence modes are TEMs. It is well-known that the streamer-type TEM turbulence is highly effective in transporting plasma energy along the radial streamers for electrostatic potential perturbations on the order $10^{-2}$ relative to the electron thermal energy [10]. At the same time, evidence exists that blobby turbulence may not be effective in the radial transport of plasma energy and that the heat-flux spreading seen in present devices is mostly from the ion neoclassical orbit effect [3, 6]. Details of the electron and ion transport in blobby turbulence are the subject of an on-going study. We note here that due to the high drift frequency in the H-mode edge, $\omega_*\sim v_{th}(\rho/L)$ with a short gradient scale length $L$, the weakly collisional trapped electron modes can easily be triggered at higher electron kinetic energies – according to the resonance relation $\omega_*\sim U_{precess}\sim v(\rho/R)(B_0/B_P)$ – around the magnetic separatrix if the effective electron collision frequency is low $\nu_{e*} \lesssim 1$ and the local ExB-flow shearing rate is low. Using the XGC1 simulation parameters, we find $\nu_{e*}(\psi_{99}, q_{95})\simeq 0.9$ for the ITER 12.5MA edge and $\nu_{e*}(\psi_{99}, q_{95})\simeq 0.95$ for the ITER 15MA edge, where $\nu_{e*}(\psi_{99}, q_{95})$ is defined using the plasma density and temperature at $\psi_{99}$, but the safety factor q is measured at $\psi_{95}$. We also find that $\nu_{e*}(\psi_{99}, q_{95})$ for the ITER 5MA edge is similarly low, indicating that the low electron collisionality is not a sufficient condition for the occurrence of a wide $\lambda_q^{XGC}$, but only a necessary condition (requiring a weak E×B-flow shearing rate also).

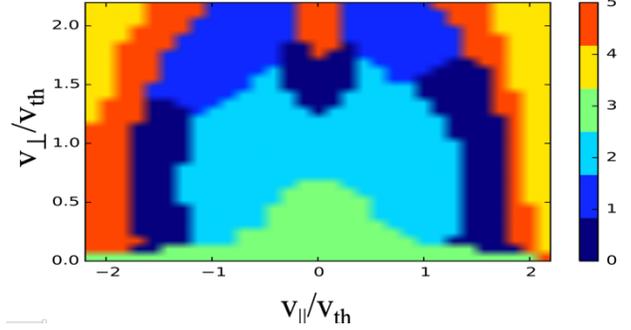

*Fig. 10. Electron response correlation in velocity space to the edge turbulence just outside of the separatrix surface, found from unsupervised machine-learning. K-Means Clustering method, specifying only six groups, is used using APACHE Spark. It can be seen that the trapped electrons around the energy band $[(v_{\perp}^2 + v_{\parallel}^2)/v_e^2]^{1/2} \sim 2$ respond to the edge turbulence in a correlated way. Reproduced with permission from IEEE Transactions on Plasma Science **48**, 2661 (2020). Copyright 2020 IEEE.*

In fact, together with the low electron collisionality, a weak E×B-flow shearing rate across the separatrix surface in the high current ITER edge is observed in XGC1, while a strong E×B shearing rate is always observed in XGC1 – and in the laboratory experiments – in the edge of present tokamaks. Figure 11(a) depicts the mean electrostatic potential profile in the pedestal and across the separatrix of the 15MA ITER plasma and, for comparison, the equivalent for the JET 4.5 MA plasma in Fig. 11(b). Vertical axes are approximately scaled to be proportional to the pedestal temperature for each plasma: 5 keV for the 15MA ITER pedestal and 1.75keV for the 4.5 MA JET pedestal. A large difference in the E×B-flow shearing rates across the magnetic separatrix can be easily implied from these figures. The actual E×B-flow shearing rate across $\Psi_N=1$ (normalized to diamagnetic frequency at $k_{\perp} = 1/\rho_{i,pol}$) is in fact compared in Fig. 12 for the JET 4.5MA and 15MA ITER discharges, together with the 1.5MA DIII-D case. We comment here in passing that the zonal flow oscillations are more noticeable in the 15MA ITER edge, which will be further subject for future study.

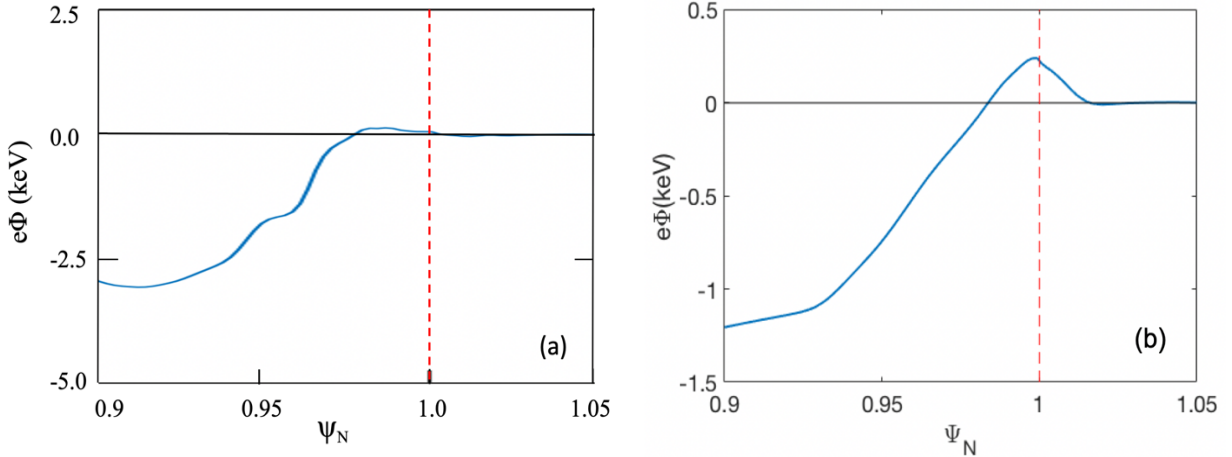

*Fig. 11. Structure of the mean, flux-surface-average electrostatic potential e<Φ> across the pedestal and magnetic separatrix from XGC1 as function of the normalized radial coordinate $\psi_N$ (poloidal magnetic flux) for (a) the 15MA ITER edge where a relatively small variation across the separatrix surface can be seen that leads to a weak E×B flow and hence its shearing rate and (b) the 4.5MA JET edge where the relatively strong e<Φ> variation across the separatrix is a typical phenomenon seen in present-day tokamaks.*

For reference, we show in Fig. 13 the plasma density and temperature profile inputs used in the XGC1 simulation of the 15MA ITER plasma which produced Fig. 11(a). The blue lines represent the electron density ($n_e$) and temperature ($T_e$) input profiles initially tried in XGC1, supplied from JINTRAC integrated modeling of a 15MA ITER deuterium-plasma. The modelled ion temperature ($T_i$) profile is not shown, but is similar to $T_e$, with its value somewhat higher (lower) than $T_i$ in the core (pedestal) region. As explained in Ref. [6] and earlier in this paper, XGC1 found that the ion-scale turbulence level was too high to maintain the JINTRAC-modeled $n_e$ and $T_{e,i}$ profiles and, as a result, the plasma power flow across the separatrix and to the divertor plates was an order of magnitude higher than the edge power flow of 100MW expected in a Q=10 ITER burning plasma (50 MW additional heating, 100 MW alpha heating and 50 MW of core radiation).

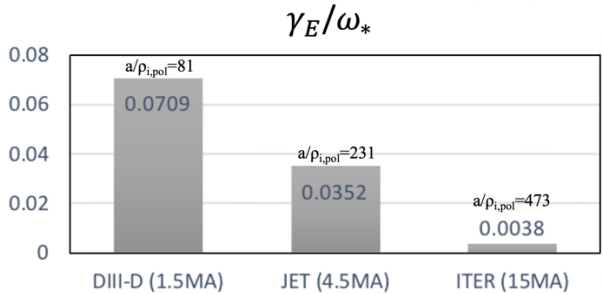

*Fig. 12. Comparison of ExB flow shearing rate $\gamma_E$ normalized to the electron diamagnetic frequency at $\Psi_N=1$ among 1.5MA DIII-D, 4.5MA JET and 15MA ITER, obtained from XGC1 (see Table 1). The diamagnetic frequency is measured at the wavelength using $B_{pol,MP}$. $a/\rho_{i,pol}$ values are 81, 231, and 473, respectively, for 1.5MA DIII-D, 4.5MA JET and 15MA ITER.*

Following the direction of XGC1's pedestal profile relaxation, we ended up with the $n_e$ pedestal shape input (red line) as shown in Fig. 13(a), and the $T_e$ and $T_i$ pedestal shapes plotted in Fig. 13(b) in red and yellow lines with an approximate power balance between the power crossing the separatrix (≈100MW) and the total power deposited onto the divertor plates (≈90MW).

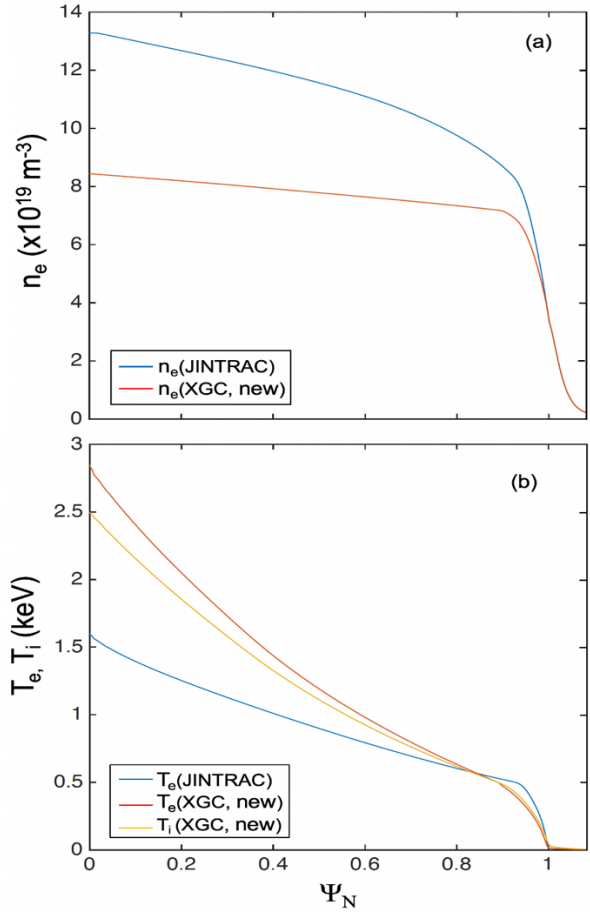

*Fig. 13. XGC1-adjusted input pedestal-SOL density in (a) and temperatures in (b), which do not evolve noticeably further at the end of the simulation. The core profiles are from artificial models and not to be trusted.*

The plasma profiles deep in the core region, manufactured to have similar electron and ion pressure as in the original JINTRAC model, are not to be trusted since the core turbulence had not yet been established by the time the XGC1 simulation was stopped. This is done to save computational time and is based on the criterion that the turbulence at the separatrix/SOL and the divertor heat flux footprint are saturated. The central plasma profiles still stay at the manufactured input level without being given a chance to evolve to a power balance. It will be an important future work to perform a much longer simulation, especially with electromagnetic turbulence, to find the self-organized plasma density and temperature values in the pedestal and central core of 15MA ITER that are consistent with the 150MW additional + alpha heating and turbulent/neoclassical transport. We also note here that: i) the outer divertor power-load was only ~25% higher than that at the inner divertor in the 15MA ITER plasma, unlike in the present tokamaks (and in fluid modelling of attached ITER burning plasmas with the SOLPS-ITER code [16]) where XGC1 finds that the outer divertor power-load is almost twice higher; ii) the divertor heat-flux width on outer divertor target is not well correlated with the plasma decay length in the near-SOL along the outer midplane (the so-called density SOL width). The cause of observation i) is an equilibrated ion power deposition between inboard and outboard divertor plates, while the inboard electron power load is only about half of the outboard power load as observed in the present tokamak simulations. Preliminary results on the parameter dependence of the out/in divertor power deposition asymmetry will be presented later in this section. The observation ii) indicates that the plasma energy crossing below the outboard midplane may be more important than the flux-tube connection effect between the outer divertor and outboard midplane. These topics are not well studied and are as yet inconclusive yet. They require more careful study in the future.

We caution here that the flux-surface-averaged mean electrostatic potential <Φ> in the far SOL shown in Fig. 11 may not be physically meaningful. Only the shape of <Φ> in the near-SOL – and radially inward – needs to be considered physical, with an unknown additive constant. First of all, what is solved in the gyrokinetic Poisson equation is not the absolute electrostatic potential value itself, but the first and the second derivatives of the electrostatic potential under a given boundary condition. Secondly, we use an artificial Dirichlet boundary condition (<Φ>=0) at the flux surface where the field lines connect to a material surface. In the case of the 15 MA ITER plasma, the contact of the plasma with the first wall occurs at the low field side. In other words, our axisymmetric electrostatic potential in the SOL is non-zero only in the region where the field lines intersect the inner and outer divertor plates without being intercepted by the first wall. Since the first wall surface touches the edge plasma only in certain small areas, large areas of the flux surface are filled with plasma which continues into the first wall shadow. In the real tokamak plasmas, this flux surface may have a mean positive $<\Phi>$ value relative to the limiter surface on the order of electron thermal energy. The reason for using an artificial $<\Phi>=0$ Dirichlet boundary condition before reaching the real material wall in these simulations is that when the particle number density becomes too low in the limiter/first-wall shadow, our axisymmetric Poisson solver sometimes does not give a converged solution. As a consequence of these assumptions in the far-SOL, we can only discuss the mean radial electric field and its shearing rate in the near-SOL, across the magnetic separatrix, and inward into pedestal in Fig. 11.

There could also be a question of how the steep H-mode pedestal gradient can be supported in the radial force balance equation at $\Psi_N>0.98$ of the full-current ITER edge plasma where the radial electric field is small, as shown in Fig. 11(a). For the sake of argument, we use the radial force balance equation (1) derived for the closed flux surface, even though it may not be highly accurate across the separatrix surface. XGC1 finds that the plasma gradient across the magnetic separatrix ($0.98 < \Psi_N < 1.01$) is maintained by the local co-current parallel/toroidal flow across the magnetic separatrix [see Eq. (2) for a simpler equation]. We demonstrate this phenomenon in Fig. 14 by showing two representative forces across the $\Psi_N=1$ surface: the radial force term from co-current toroidal flow ($\approx<v_\parallel>$, green line) which is of the same order of magnitude and opposite to the radial density gradient force (dashed line). Other terms are less significant and are not shown in the figure. The physics origin for this phenomenon is the X-point orbit-loss driven $E_r$ and toroidal torque [31]. The neoclassical dielectric/polarization effect [31, 32] and the collisional damping of poloidal plasma rotation in a tokamak plasma [17, 33] can easily suppress the weak radial electric field, but the weak toroidal viscosity cannot easily suppress the toroidal rotation. Without the radial electric field opposing the X-point orbit-loss driven toroidal flow, the toroidal flow can replace the role of the radial electric field. A discussion of the physics of kinetic co-current edge momentum generation across the magnetic separatrix by X-point orbit loss torque can be found in Ref. [31, 34].

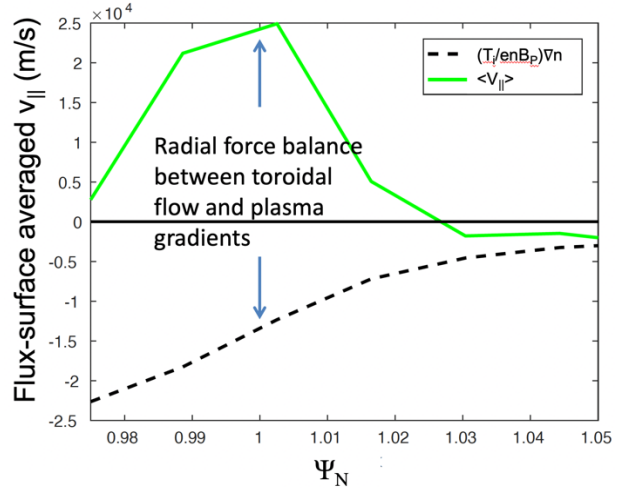

Fig. 14. Two representative terms across the separatrix surface in the fluid radial force balance equation obtained from XGC in the 15MA ITER simulation.

The spatial turbulence pattern of the 10MA steady-state ITER edge plasma is of special interest, since it shows only a partial enhancement of $\lambda_q^{XGC}$ compared with the expected experimental scaling value. It can be seen from Fig. 15 that the temperature-normalized electrostatic potential fluctuation across the outboard-midplane magnetic separatrix is a mixture of blobs (isolated structures at high amplitude, red and blue) and streamers (connected structures at low amplitude). The streamer feature has not been seen in the XGC1 simulations of present tokamaks, where only the blob feature has been observed. The partial enhancement of $\lambda_q^{XGC}$ in the 10MA ITER edge appears to be from the low amplitude streamers, which are known to be highly effective carriers of heat from core-region turbulence studies [10], as explained earlier. This is valuable information. The large enhancement of $\lambda_q^{XGC}$ in the 15MA or 12.5MA ITER plasma is not from a sudden physics bifurcation, but is a gradual effect occurring as a result of the transition from blobs to streamer transport. An explicit transport mechanism study of kinetic electron and ion particles as they pass through the blobs and streamers in the open-field line region under parallel streaming and perpendicular drift motions is presently underway using an in-situ data management technology. It will be reported in the near future.

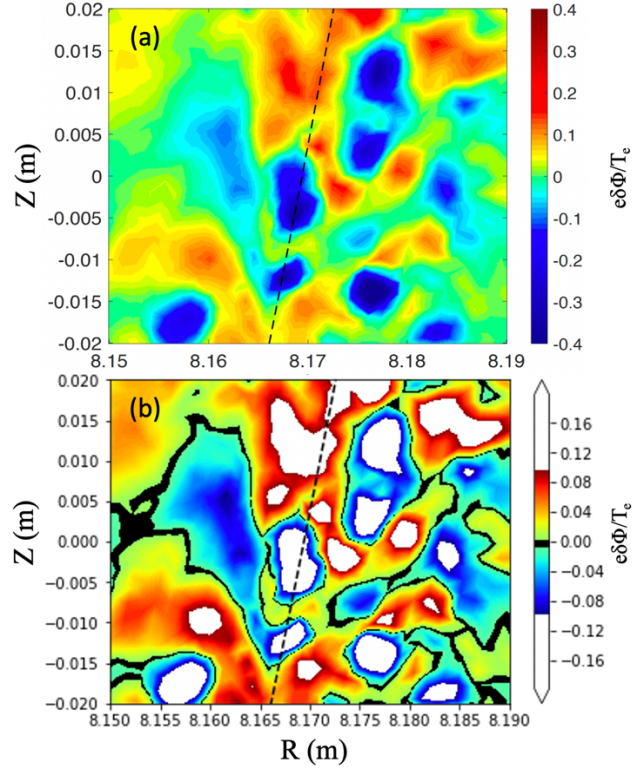

Fig. 15. Turbulence structure across the outboard midplane separatrix (dashed line) of the 10MA ITER plasma. (a) A mixture of blobs (dark blue and red at high amplitude) and streamers (most noticeable in the pale greenish yellow color) can be seen. (b) For an enhanced coloring of the mixture between isolated blobs and connected streamers, blob tops at $|e\delta\Phi/T_e| \geq 0.1$ are colored white and the negligibly small fluctuation amplitude at $|e\delta\Phi/T_e| \leq 0.003$ is colored black. Pale green and blue colors highlight the streamer range amplitude, showing connected structures across the separatrix surface.

Another noteworthy observation we have made from the gyrokinetic ITER simulations is the dependence of the power deposition ratio between the outer and inner divertor plates on the new scaling parameter $B_{pol,MP}\, a/\rho_{i,p}$ used in the machine learning approach. As shown in Fig. 16(a), the out/in power ratio decreases as $B_{pol,MP}\, a/\rho_{i,p}$ increases from the 5MA plasma to the 10MA and 15MA plasmas. At 5MA, the out/in ratio of ~1.7 is similar to the present tokamak values. At 15MA, the ratio decreases to 1.25. The peculiar 12.5MA plasma (star mark), though, shows an irregular behavior compared to the other cases. This could mean that the reduction amount of the outer/inner power deposition ratio in the ITER 15MA could be subject to some unknown effects that need to be studied.

Fig. 16(b) depicts the same graph as in Fig. 16(a), but now as a function of $a/\rho_{i,p}$. The same trend is found, meaning that the out/in power deposition ratio behavior cannot be definitely identified as due to the enhanced $B_{pol,MP}(T)\, a/\rho_{i,p\ value}$ or the enhanced $a/\rho_{i,p}$ value. It appears that the reduction in the inner/outer divertor power deposition ratio from 5MA, to 10MA and to 15MA is related to the co-current directional parallel plasma flow, thus positive poloidal flow, across the separatrix surface (see green line in Fig. 13) which could bring more plasma power to the inner divertor plates. In common with several other detailed phenomena observed from the simulations, further work is required to provide a more definitive answer to this question.

## VII. Summary and Discussion

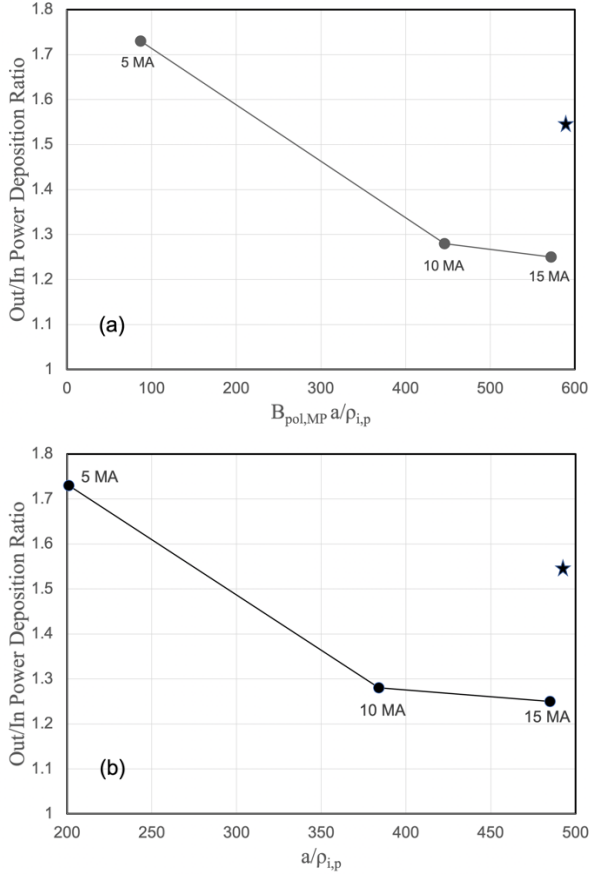

Fig. 16. (a) Out/in divertor power deposition asymmetry factor $P_{asym}$ decreases linearly with the new parameter $B_{pol,MP}(T)\, a/\rho_{i,p}$. The star mark represents the peculiar 12.5MA case. (b) $P_{asym}$ exhibits a similar decrease trend with $a/\rho_{i,p}$.

The XGC1 gyrokinetic particle-in-cell code in the electrostatic mode, with which the predictive divertor heat-flux width simulations have reproduced the experimentally measured $\lambda_q$ from the three US major tokamaks in the attached (inter-ELM H-mode) low recycling divertor regime, has reported a much wider divertor heat-flux width, $\lambda_q^{XGC}$ for the full-current (15MA) ITER model plasma than expected on the basis of the experimental scaling [6]. Several new simulations are performed to answer some essential questions following the previous report. How would XGC1 predict $\lambda_q^{XGC}$ on the highest current JET plasma, which has $B_{pol,MP}(=0.89T)$ only 26% lower than the $B_{pol,MP}(=1.21T)$ of the full current ITER? This question was especially worthy to answer because the old JET experimental data showed some broadening-like data points even at lower $B_{po,MP}$ values, as can be seen in Fig. 1 or in the Eich-scaling reports [1, 2] (red circular dots). Were these old JET data from inaccurate experimental measurement or from real physics? Our simulation predicts that the highest-$B_{pol,MP}$ JET discharge has $\lambda_q^{XGC}$ that is consistent with the Eich scaling (red open circle in Fig. 1). This result could suggest a possibility for a $\lambda_q^{XGC}$ bifurcation between $B_{pol,MP} = 0.89T$ of JET and the $B_{pol,MP} = 1.21T$ of the 15MA ITER discharge.

A more significant question then arises. In a C-Mod experiment, $B_{pol,MP}$ was raised to the level of the full-current ITER and it was found that $\lambda_q^{Exp}$ still follows the Eich formulas. An XGC1 simulation performed and agreed with the experimental finding (see the black open star symbol at the far-right bottom of Fig. 1), giving rise to double valued solutions if $B_{pol,MP}$ is the sole parameter in $\lambda_q^{Eich(14)}$. This questioned the existence of a bifurcation of $\lambda_q^{XGC}$ with $B_{pol,MP}$ and suggested a hidden parameter outside of the macroscopic parameter set used in Refs. [1-3].

A supervised machine-learning tool is applied to all the $\lambda_q^{XGC}$ data points (together with the corresponding experimental data points $\lambda_q^{Exp}$) obtained for the existing tokamaks and the full-current 15 MA Q = 10 ITER plasma, with a feature engineering of adding the physics-based kinetic parameter $a/\rho_{i,pol}$ to $B_{pol,MP}$. The result, shown in Fig. 3, is a new simple formula for $\lambda_q^{XGC}$ that reduces to $\lambda_q^{Eich(14)}$ in the present tokamak regime including the highest current C-Mod case, that reproduces the full-current ITER result, and that is physically meaningful. The new additional simplest dependence parameter is found to be $B_{po,MP}(a/\rho_{i,pol})$, combination of the neoclassical E×B-flow shearing rate parameter $\rho_{i,pol}/a$ and the ion orbit width parameter $1/B_{po,MP}$.

Tests of the new formula are performed using a 5MA H-mode ITER plasma which has a $B_{po,MP}$ (a/$\rho_{i,pol}$) value similar to that in existing tokamaks, a 12.5MA Q = 5 long pulse ITER plasma with $B_{po,MP}$ (a/$\rho_{i,pol}$) slightly greater than the full-current ITER plasma, and a 10MA Q = 5 steady-state ITER plasma which has $B_{po,MP}$ (a/$\rho_{i,pol}$) in the gap between the highest current JET and the full current 15MA ITER. The new simplest formula well survives against these tests, as depicted in Fig.5. Other, more complicated formulas suggested by the machine learning program did not do well against the 10MA ITER test, which lies deep in the gap region between the JET and the 15MA ITER points in the new parameter space.

In an effort to study the new physics that leads to the $\lambda_q^{XGC}$ broadening in the full-current ITER and that is consistent with the new parameter, three independent data analyses are performed. The study identifies the new physics to be weakly-collisional, trapped-electron driven turbulence, gradually dominating over the blobby turbulence as the new parameter $B_{po,MP}$ (a/$\rho_{i,pol}$) increases.

We comment here that the main difference in the present gyrokinetic simulation results from the low divertor pressure case in the recent 15MA ITER result of Kaveeva et al. [16], which used SOLPS-ITER code with an assumed anomalous electron thermal diffusivity of 1 m$^2$/s in SOL are: i) much smaller value of the E×B-flow shear across the separatrix, ii) ~2X wider outer divertor heat-load width, iii) weaker outboard/inboard power load ratio, and iv) smaller effective heat diffusivity at ≈ 0.2m$^2$/s (an averaged value across the separatrix surface 0.98≤ $\Psi_N$ ≤ 1.02). The physics relationship between the ~2X wider outer divertor heat-load width and the eventual relaxation to ~2X wider edge pedestal width has not been established from the present gyrokinetic simulations. As stated in our previous report [6], $\lambda_q^{XGC}$ saturates before the ~2X relaxation of the pedestal width is reached. The above-quoted effective radial diffusion coefficient is only a ball-park number. Radial plasma fluxes fluctuate significantly along the field line depending upon the space-time varying turbulence structure and, thus, a "flux-surface-averaging" is employed to obtain a statistically accurate value in a core-region plasma. In the open field region and across the separatrix surface, the survival time of an individual particle motion is short due to divertor-plate intersection and atomic physics, hence the "flux-surface-averaging" is limited and yields a higher statistical error. An advanced data analysis technique is under development to resolve this issue, by accurately following the individual particle motions in the turbulent field while obtaining statistical transport information, in a similar way to the transport measurement used in stochastic systems (see Equations 10 and 11 in Ref. [35] and the quoted references therein).

We note here that the present simulation is electrostatic. Even though the electrostatic XGC has reproduced $\lambda_q$ in the present tokamaks, the effect of the electromagnetic turbulence on $\lambda_q^{XGC}$ in the high Q ITER edge is of interest. Present studies are conducted under the low-recycling attached divertor conditions, corresponding to the condition relevant to Refs. [1-3]. ITER will have to operate in the semi-detached or detached divertor regimes for high Q plasmas. These subjects, and others, are left for future study. In addition, a way to test the new formula in the present experiments is of interest. This may require finding or creating a plasma with $\nu_{e*} \lesssim 1$ and a low-sheared E×B flow near the magnetic separatrix surface.

A shortfall not mentioned in the main text is the lack of a systematic validation metric [36] from the XGC1 simulation results due to the small number of extreme-scale simulations and highly limited availability of the experimental primacy hierarchy data in the edge plasma. Systematic validation of limited number, extreme-scale simulations is an active research topic in the uncertainty quantification community.

**Acknowledgement**

We acknowledge helpful discussion with M. Romanelli, T. Eich and R. Goldston in the early phase of the study. We thank R. Maingi, J.-W. Ahn, T. Gray, B. LaBombard , T. Leonard, M. Makowski and J. Terry for their contribution to the original paper [6] with experimental data from three US tokamaks and valuable discussions. Without their help, this work could not have been launched.

The research is mostly supported by the US DOE Office of Science SciDAC program, both from ASCR and FES, through the award the SciDAC-4 Partnership Center for High-fidelity Boundary Plasma Simulation (HBPS) under the contract No. DEAC02-09CH11466 with Princeton University for Princeton Plasma Physics Laboratory. Computing resource is provided by OLCF and ALCF via INCITE program. The simulations of ITER plasmas have been carried out as part of the coordinated work programme of the Pedestal Confinement and Stability Modelling ITER Scientist Fellows group. The data that support the findings of this study are available from the corresponding author upon reasonable request.

Disclaimer: ITER is the Nuclear Facility INB no. 174. The views and opinions expressed herein do not necessarily reflect those of the ITER Organization.